# Fundamental Physics with a State-of-the-Art Optical Clock in Space

Andrei Derevianko[1], Kurt Gibble[2], Leo Hollberg[3], Nathan R. Newbury[4], Chris Oates[4], Marianna S. Safronova[5], Laura C. Sinclair[4], and Nan Yu[6]

[1]*Department of Physics, University of Nevada, Reno, Nevada 89557, USA*
[2]*Department of Physics, The Pennsylvania State University, University Park, Pennsylvania 16802, USA*
[3]*Department of Physics, HEPL, Stanford University, 452 Lomita Mall, Stanford, California 94305 USA*
[4]*National Institute of Standards and Technology, 325 Broadway, Boulder, CO 80305, USA*
[5]*Department of Physics and Astronomy, University of Delaware, Newark, Delaware 19716, USA*
[6]*Jet Propulsion Laboratory, California Institute of Technology, Pasadena, CA 91109, USA*
[Dated December 20, 2021]

Recent advances in optical atomic clocks and optical time transfer have enabled new possibilities in precision metrology for both tests of fundamental physics and timing applications. Here we describe a space mission concept that would place a state-of-the-art optical atomic clock in an eccentric orbit around Earth. A high stability laser link would connect the relative time, range, and velocity of the orbiting spacecraft to earthbound stations. The primary goal for this mission would be to test the gravitational redshift, a classical test of general relativity, with a sensitivity 30,000 times beyond current limits. Additional science objectives include other tests of relativity, enhanced searches for dark matter and drifts in fundamental constants, and establishing a high accuracy international time/geodesic reference.

## 1. Introduction

### A. Optical Atomic Clocks – Progress and the Need for Space

Time is an omnipresent concept in modern society and plays a central role in the foundations of physics and our understanding of the cosmos. Atomic clocks keep international time and their quantum measurements are, by orders of magnitude, the most accurate measurements of any physical observable. Figure 1 shows the dramatic improvement of the accuracy of atomic clocks over the last 20 years, far faster than in the last half of the 20[th] century. This extremely rapid improvement of clocks at optical frequencies resulted from advancements in the manipulation of atomic quantum systems, as well as advanced techniques in laser cooling/trapping and the development of stabilized fs-laser frequency combs in 2000, which enabled the reliable counting of the ticks of the cycles of laser light at $10^{15}$ cycles per second (Abdel-Hafiz et al. 2019; Ludlow et al. 2015; Sanner et al. 2019). The most accurate versions of these optical clocks use trapped quantum absorbers (either ions or lattice-confined neutral atoms) and now have fractional frequency uncertainties approaching or exceeding 1 part in $10^{18}$ (Bothwell et al. 2019; Brewer et al. 2019; Godun et al. 2014; Huntemann et al. 2016; McGrew et al. 2018; Takamoto et al. 2020), which is enabling transformational advances in areas including navigation and ultra-precise ranging (Mehlstäubler et al. 2018), searches for dark matter (Kennedy et al. 2020), searches for violation of Lorentz invariance (Sanner et al. 2019), searches for variation of fundamental constants (Lange et al. 2021), and stringent tests of general relativity (Takamoto et al. 2020). Future atomic clocks are also proposed for gravitational wave detection (Ebisuzaki et al. 2019; Kolkowitz et al. 2016). Indeed, with no foreseeable barriers to continued improvement along the recent trend in Figure 1, clock accuracies could reach $10^{-19}$ in 2027, $10^{-20}$ by 2034, and $10^{-21}$ in two decades, at which point gravitational waves could be directly observed by measuring length changes directly with the optical clocks. While specific large advances are not possible to predict, dramatic improvements using quantum entanglement are anticipated (Braverman et al. 2019), and there is no known fundamental limitation in sight to the ultimate achievable accuracy of atomic clocks.

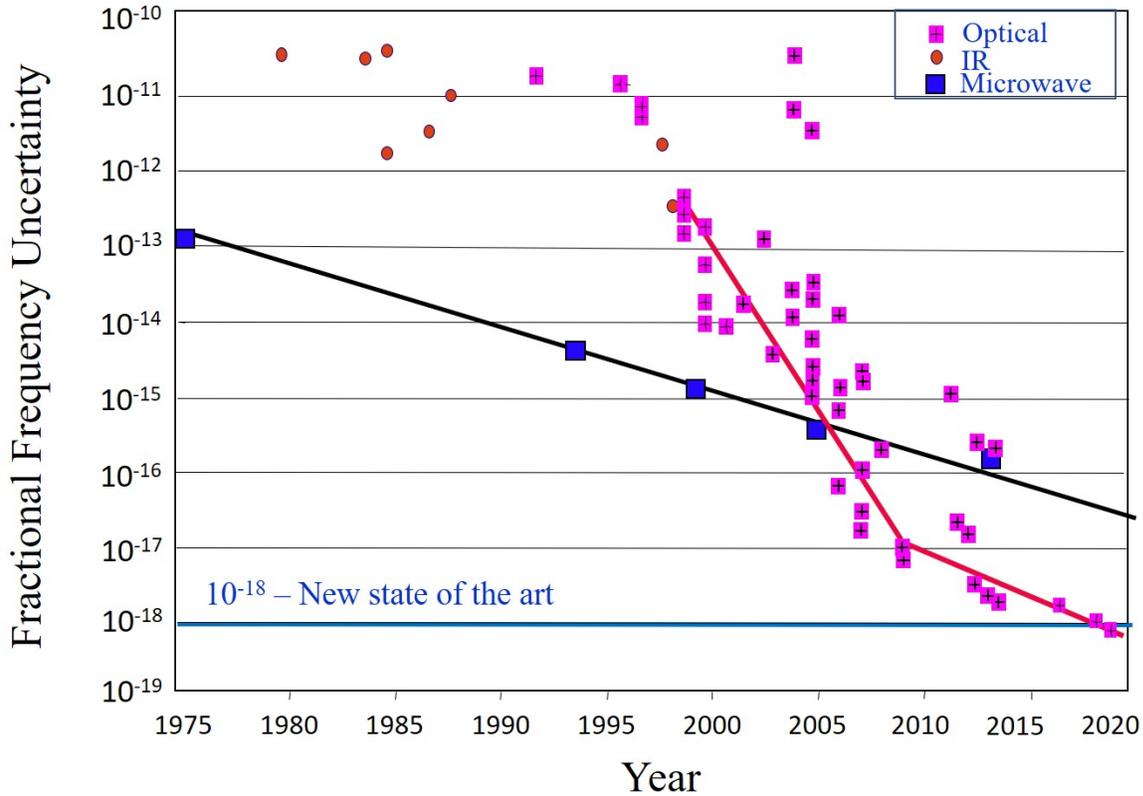

*Figure 1. Historical accuracy of atomic clocks. Since 2000, the uncertainty of optical atomic clocks have rapidly improved to, and beyond, 1 part in $10^{18}$.*

In general relativity (GR), the tick rate of time is no longer universal, but slows in the presence of massive bodies. With further advances in clock accuracies and stabilities, measurements of time on the surface of the earth will soon be limited by the instability of time itself due to gravitational fluctuations, for example from tides and seismic noise. A straightforward solution is to locate one or more clocks in orbits around the Earth, thereby avoiding Earth's tidal motion/gravitational noise and reducing the sensitivity to Earth's gravity for medium-Earth (MEO) and high Earth (HEO) orbits. Such an orbiting platform provides a low-noise environment that can enable atomic clocks to perform at the nineteenth digit and beyond. As a result, anticipated improvements in clock performance could be used in a variety of applications, including dramatically advancing tests of fundamental physics. We note that there is a strong synergy between the technology and the underlying measurements of optical atomic lattice clocks and the atom interferometers being proposed for gravitational wave (GW) observations at mid-band frequencies, complementary to the eLISA and LIGO observatories (Abe et al. 2021;

Badurina et al. 2020; Canuel et al. 2018; El-Neaj et al. 2020; Hogan & Kasevich 2016; Loriani et al. 2019; Tino et al. 2019; Tino & Vetrano 2011; Yu & Tinto 2011; Zhan et al. 2020).

These motivations have led to a number of international projects on orbiting clocks. A laser-cooled microwave clock, CACES, operated on the Chinese Tiangong-2 space station (Liu et al. 2018) and the ESA project ACES with a cold atom clock is scheduled to launch in the coming years (Cacciapuoti et al. 2020; Savalle et al. 2019a). For over a decade ESA has been developing optical clocks for space as part of their ISOC program (Schiller et al. 2012). The German Aerospace Center (DLR) is developing a combined Iodine clock and frequency comb for the ISS (COMPASSO), and the Chinese Space Agency aims to demonstrate an optical lattice clock in orbit on their next generation space station (Klotz). In contrast to the more modest clocks/links used in these projects, we propose below a mission that aims to deploy an optical clock and link with state-of-the-art performance in a modulated spacetime/gravity environment that will yield an ultra-sensitive space-time probe for fundamental physics with uncertainties reduced dramatically below current limits.

**B. Testing General Relativity and the Standard Model of Physics with atomic clocks**

In terms of fundamental physics, the Standard Model and General Relativity describe a vast array of physical phenomena and have passed nearly every precision test to date (Muon g−2 Collaboration et al. 2021). However, these theories cannot coexist in their present form to provide a quantum description of gravity and are unable to account for key phenomena, such as dark energy, dark matter, the matter/anti-matter imbalance, and the unique direction of time. Indeed, arguably the most significant problems facing physics today are connected to this conundrum. Thus, there is a strong motivation to find new theories or extend existing ones to address these gaps. However, experimental confirmation of such extensions has thus far proven elusive. New physics can appear either at extremely high energy scales, ("the energy frontier") where new particles can appear as distinct resonances, or at low energies where colliders are blind to new physics with feeble couplings to standard particles due to background rejection ("the high precision frontier") (Safronova et al. 2018). At low energies, the competing theories with "new physics" have indicated possible places where the Standard Model of particle physics might fail, including the three components of the Einstein Equivalence Principle (EEP): (1) weak equivalence principle (WEP) and, therefore, universality of free fall, (2) local Lorentz invariance (LLI), and (3) the local position invariance (LPI) (Will 2014). As a result, over the past decades there has been a notable increase in proposed and experimental tests of these basic theories that look for new physics or gaps in our existing physics (Delva, Hees & Wolf 2017; Safronova et al. 2018). Many of these tests are based on time/frequency metrology, which is a direct consequence of the capability to measure time nearly a billion times more precisely than other base SI units. As a result, atomic clocks are presently one of the tools of choice to pursue more stringent tests of fundamental theories.

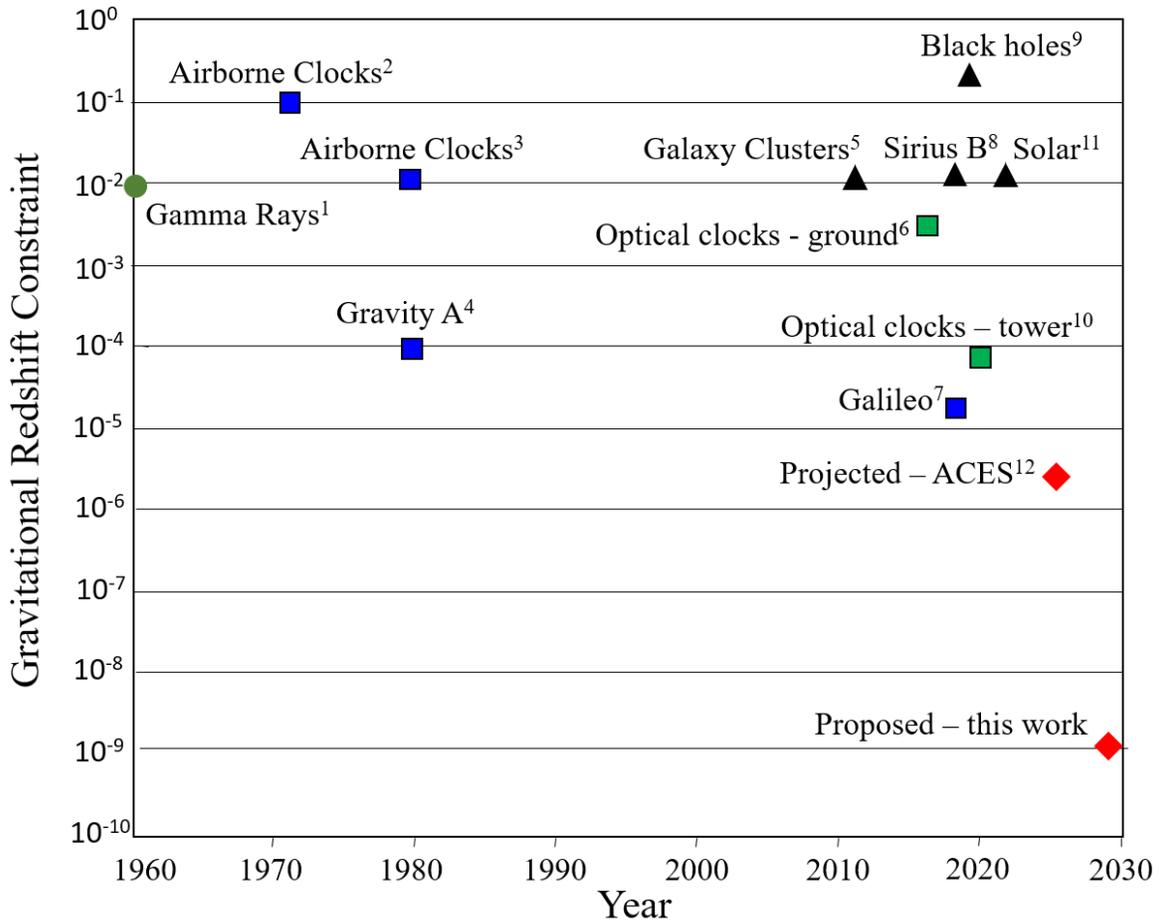

*Figure 2: Constraints on the gravitational redshift parameter α over the past six decades. Green squares represent Earth-based measurements, blue squares represent space-based measurements, black triangles represent astronomical measurements, and red squares represent proposed space-based experiments. Combining the recent rapid advances of optical clocks with a large change in gravitational potential enables a dramatically improved test of General Relativity.* [1]*(Pound & Snider 1965)* [2]*(Hafele & Keating 1972)* [3]*(Alley 1979)* [4]*(Vessot et al. 1980)* [5]*(Wojtak, Hansen & Hjorth 2011)* [6]*(Takano et al. 2016)* [7]*(Delva et al. 2018)* [8]*(Joyce et al. 2018)* [9]*(Do et al. 2019)* [10]*(Takamoto et al. 2020)* [11]*(González Hernández et al. 2020)* [12]*(Cacciapuoti et al. 2020; Savalle et al. 2019a).*

In particular, the rapid advancement of state-of-the-art optical clocks shown in Figure 1 has enabled new possibilities for using such clocks in space to dramatically advance tests of fundamental physics. As an example, we show in Figure 2 a representative measurement history of one of the classic tests of fundamental physics, the gravitational redshift, a direct consequence of Local Position Invariance of EEP [see (Takamoto et al. 2020) and references therein]. For a difference in gravitational potential, $\Delta U$, the fractional frequency shift $\frac{\Delta \nu}{\nu} = (1 + \alpha)\frac{\Delta U}{c^2}$, represents a measure of the deviation of a given measurement from the predictions of GR, parameterized by the violation $\alpha$. Bounds on α have been tightened through the years with ground- and space-based measurements. We note from Figure 2 that the high precision atomic clock measurements provide more stringent constraints on the gravitational redshift than astrophysical measurements (black triangles), even though they are done in the relatively weak gravity field of Earth. (In fact, the redshift measurements of stars may be more useful in determining the mass of the star by taking advantage of the low uncertainty from the clock measurements of the red shift (Do et al. 2019).) We note also that since the initial tower-based measurement in 1960 (Pound & Snider 1965), tighter constraints have come via experiments that put clocks in a space environment (Delva et al. 2018; Herrmann et al. 2018; Vessot et al. 1980). But perhaps the most significant takeaway from Figure 2 is that, because such experiments are difficult and expensive, the overall progress in the reduction of the constraints on α has been relatively slow, with less than a factor of 10 improvement over 40 years. For comparison, Figure 1 shows the reduction of atomic clock uncertainties by a factor of almost a million over a similar period.

There are also differential redshift experiments, referred to as "null" experiments in (Will 2014), where the relative rates of two different clocks are measured throughout the year (such tests were also conducted with Dy (Leefer et al. 2013). Since clocks based on different atomic transitions have different sensitivities to the variation of fundamental constants, by measuring the relative variance of frequencies in the gravitational potential of the Sun due to the eccentricity of the Earth orbit ($\Delta U/c^2 = 1.65 \times 10^{-10}$, where $c$ is the speed of light), these experiments essentially test the variation of fundamental constants due to changes in a gravitational potential (Safronova et al. 2018). The Yb$^+$ electric quadrupole (E2) and electric octupole (E3) pair of clock transitions has the highest sensitivity to the variation of α of current clocks. The most recent comparisons of these Yb$^+$ E2 and E3 transitions and Cs clocks have set constraints of $(c^2/\alpha_{FS})(d\alpha_{FS}/dU) = 14(11) \times 10^{-9}$ and $(c^2/\mu)(d\mu/dU) = 14(11) \times 10^{-9}$, where $\alpha_{FS}$ is the fine-structure constant and $\mu$ is the proton-to electron mass ratio (Lange et al. 2021).

Figures 1 and 2 thus stimulate the question of how we can significantly advance tests of GR and other fundamental theories? Given that the sensitivities of many fundamental tests, such as the redshift, are proportional to the clock performance, we now have a golden opportunity: by strategically deploying $10^{-18}$ clocks in space, we can leverage the recent advances in clock performance and supporting breakthroughs in time transfer to enable tests of GR and the Standard Model with a leap in sensitivity, thereby significantly tightening constraints on key parameters or discovering EEP violation. Many of these parameters are connected to Planck scale physics, and possibly new fields and forces, complementary to those explored by high energy accelerator physics and astrophysics observatories. We note that on the one hand, it is challenging to further increase the energy scales of accelerators and, on the other hand, observational cosmological investigations of dark matter and dark energy will benefit from the major science missions Euclid and the Roman Telescope (WFIRST). Precision measurements enabled by clocks and frequency control can probe the yet unconstrained parameters of possible new physics, thereby providing prospects to elucidate science mysteries and discover new physics.

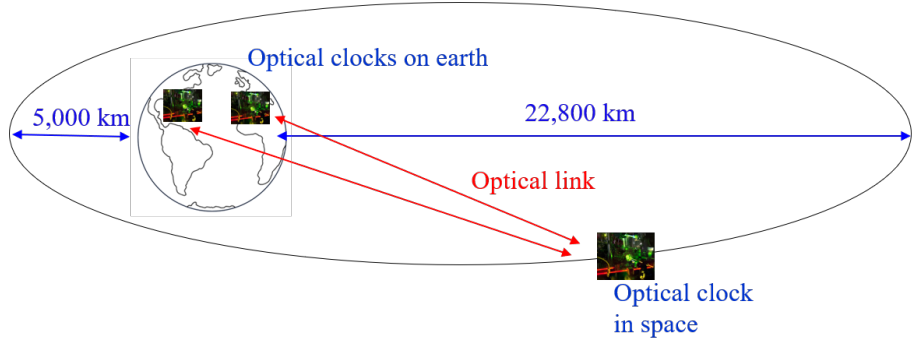

*Figure 3. Schematic of the proposed mission to test **F**undamental physics with an **O**ptical **C**lock **O**rbiting in **S**pace (FOCOS). A high-performance optical clock in an elliptical orbit around Earth is connected to ground optical clocks through a high stability optical link. Clock comparisons will enable tests of fundamental physics and provide an international clock reference for timing/geodetic applications.*

Here we consider a state-of-the-art optical atomic clock in an eccentric orbit around Earth, which provides a relatively large variation of the gravitational potential (Fig. 3), to achieve multiple fundamental science goals. We propose connecting a single orbiting clock with earthbound clocks through high performance optical links to enable high stability earth/space clock comparisons. The modulation of the gravitational potential in turn modulates the frequency of the space clock with an accurately known period. This modulation is measured through the optical link, which enables evaluation of gravitational frequency shifts at uncertainty levels below the absolute accuracy of the clocks being compared and makes it easier to separate gravitational effects from possible systematic drifts in the clocks and the links. As a result, the space clock system, with a planned fractional frequency instability of $1 \times 10^{-16}\tau^{-1/2}$ and fractional inaccuracy of $1 \times 10^{-18}$, where $\tau$ is the measurement averaging time in seconds, will enable a measurement of the gravitational redshift with a sensitivity 30,000 times higher than previously achieved (see Fig. 2). We note that variations of such experiments with optical clocks in space have been proposed, e.g. (Altschul et al. 2015; Litvinov & Pilipenko 2021).

A clock orbiting in space near Earth will also enable tests of local Lorentz invariance and searches for hypothetical ultralight fields. Taken together, these measurements will contribute significantly to our understanding of the basic framework of the universe and will help constrain or support new theories of spacetime/gravity that attempt to explain physical phenomena (dark energy, dark matter, quantum gravity) not presently accounted for in existing theories. Moreover, a space-based clock will provide an ultimate timing/geodesic reference frame that is freed from the noisy gravitational environment of the earth's surface that is expected to contribute significantly to clock uncertainty budgets in the $10^{-19}$ decade. Such an orbiting reference could be used to connect widely separated earthbound optical atomic clocks to create a global network to perform fundamental physics tests at unprecedented levels, such as dark-matter induced variation of fundamental constants, searches for gravity-atom orientation coupling, and searches for new physics field emission from black-hole mergers, in addition to using the GR shift for geodesy and static gravity measurements. Finally, this proposed mission, FOCOS (**F**undamental physics with an **O**ptical **C**lock **O**rbiting in **S**pace), will lay the groundwork for subsequent missions with the longer-term goal of using space-based constellations of optical clocks to search for dark matter e.g., (Roberts et al. 2017), and to observe mid-band gravity waves (Ebisuzaki et al. 2019; Kolkowitz et al. 2016; National Research Council 2011; Turyshev et al. 2007), including potentially using asteroids as test masses (Graham 2021).

In the following sections, we describe the mission science goals in more detail, mission design/requirements, and the required mission payloads, including technological choice justifications and key technological gaps.

## 2. Science Opportunities

Testing General Relativity with clocks has a history of more than 50 years. Classical General Relativity (GR) provides a geometric description of the gravitation interaction and is based on two principles (Delva et al. 2018), the Einstein Equivalence Principle (EEP) and the Einstein field equations derived from the Einstein-Hilbert action. GR has been extremely successful to date, passing all precision tests so far. However, there is an expectation that it will fail with sufficiently sensitive experiments, because post-GR formulations that enable quantum gravity formulations lead to deviations due to a characteristic length scale (Will 2014). Phenomenologically, tests of GR address three different aspects of EEP:

(i) Universality of Free Fall (UFF), i.e., acceleration is independent of body composition, which is also referred to as the Weak Equivalence Principle. The recent MICROSCOPE experiment took advantage of the quiet gravitational environment of space to compare the differential acceleration of two test masses made from different materials, with agreement at the $1 \times 10^{-14}$ level (Bergé et al. 2018), which surpassed the long-standing UFF measurement precision set by the ground pendulum experiments (Schlamminger et al. 2008) and the lunar laser ranging experiments (Murphy et al. 2012; Williams, Turyshev & Boggs 2004).

(ii) Local Lorentz Invariance (LLI) - non-gravitational physical laws are independent of velocity and orientation of the inertial reference frame. The tests of LLI are analyzed in the context of a phenomenological framework known as the Standard Model Extension (SME) (Colladay & Kostelecký 1998). The minimal SME Lagrangian contains every possible combination of the standard model fields that are not term-by-term Lorentz invariant, but maintains gauge invariance, energy–momentum conservation, and Lorentz invariance of the total action. The SME provides a valuable framework to compare the constraints from very different experiments for the same SME coefficients. We note that SME allows for separate violation of LLI by all particles, which makes it compelling to stringently verify LLI in different systems. Atomic physics tests of LLI have been carried out with atomic clocks and high-precision dysprosium spectroscopy, magnetometers, electromagnetic cavities, and quantum-information-trapped-ion technologies (Safronova et al. 2018). Atomic physics LLI tests set some of the highest bounds on the SME coefficients in the photon, electron, neutron, and proton sectors. Optical atomic clocks (Yb$^+$) produced the most stringent limits, of the order of $10^{-21}$, on Lorentz symmetry violation parameters for electron (Sanner et al. 2019). The anisotropy of the speed of light has been constrained by Michelson-Morley-type experiments. In the SME framework, this is an effect of Lorentz violation in the photon sector, which can also be probed with atomic physics experiments (as it affects the Coulomb interaction). The most recent LLI violation bounds for all sectors are listed in the 2021 edition of the Data Tables for Lorentz and CPT Violation (Kostelecky & Russell 2008).

(iii) Local Position Invariance (LPI) - non-gravitational physical laws are independent of location in time and space of a freely falling reference frame. Tests of LPI include searching for deviations from the predicted frequency shifts due to the gravitational redshift and changes in the values of fundamental constants as a function of position or time. These experiments often employ atomic clocks and have a sensitivity proportional to clock performance. The most sensitive redshift test, performed with the orbiting Galileo clocks, set a fractional redshift constraint at $< 3 \times 10^{-5}$ (Delva et al. 2018; Herrmann et al. 2018). Local position invariance also refers to position in time. If LPI is satisfied, the fundamental constants of non-gravitational physics should be constants in time. Earth-based atomic clock tests have put the most stringent limits on drifts of the fine structure constant and μ, the proton-electron mass ratio, at $1.0(1.1) \times 10^{-18}$/year and $-8(36) \times 10^{-18}$/year, respectively (Lange et al. 2021).

Within this context we give an overview of the fundamental advances possible with FOCOS:

- **Gravitational redshift test (LPI):**

The landmark Gravity Probe A experiment placed a hydrogen maser on a rocket that was then launched to 10,200 km to perform the first space-based measurement of the gravitational redshift (Vessot et al. 1980). This constrained the gravitational redshift parameter α to $\sim 1 \times 10^{-4}$. More recently, the most sensitive redshift test to date was performed by microwave atomic clocks in space (the Galileo Experiment), which was able to constrain the redshift parameter α to $< 3 \times 10^{-5}$ (Delva et al. 2018; Herrmann et al. 2018) even though the mission was not designed for this measurement. In the FOCOS mission the primary goal is to advance the state-of-the-art of this classic LPI test by a factor of 30,000 by using optical atomic clocks and optical time transfer in an optimized orbit to reach an uncertainty of $1 \times 10^{-9}$ for α. A deviation of α from zero would indicate new physics beyond our existing frameworks. We note that the GRACE, GOCE and GRACE-FO missions (along with data from geodetic satellites measured with International Laser Ranging Systems (ILRS)) provide sufficient knowledge of the gravitational potential of the Earth to enable quantitative comparisons with the experimental results from this mission.

- **Tests of local Lorentz invariance (LLI)**

The analysis of data from space-based clocks in the SME context is discussed in ref. (Bluhm et al. 2002, 2003), which describes the natural advantage of space-based experiments ability to directly access all spatial components of the basic coefficients for Lorentz and CPT violation. These results can be extended to the generally poorly constrained nonminimal sector using the methods described in (Bluhm et al. 2003).

- **Combined relativistic effects on the clock frequency and satellite orbit (LPI/LLI)**

Comparing a clock in Earth orbit with ground clocks with a fractional frequency uncertainty $\approx 1 \times 10^{-18}$ requires advancing the accuracy of calculations of relativistic effects. Measurements of the relativistic effects on the clock frequency and time delays will require a detailed knowledge of the satellite orbit and of the Earth's gravitational field. It will be important to quantify deformations of the Earth, as they change the position of the clock relative to the center of the Earth, which is the natural origin of the reference frame used for time keeping and position determination on Earth. Regarding the position of the satellite, the two-way Doppler-cancelling laser links required for this mission will give accurate range and Doppler information. Combining that with accurate knowledge of the geoid will provide a means to test these higher order relativistic corrections. This analysis must also include a relativistic-consistent treatment of reference frames (Geocentric and Barycentric), Earth, solar and lunar tidal effects, as well as non-gravitational perturbations to the orbit. Important strides in the theory of relativistic effects on clocks in orbit have been completed (Ashby 2003; Duchayne, Mercier & Wolf 2007; Linet & Teyssandier 2002; Müller et al. 2017; Müller, Soffel & Klioner 2008; Nelson 2011; Petit & Wolf 1994, 2005; Soffel & Frutos 2016; Teyssandier, Poncin-Lafitte & Linet 2008; Turyshev, Yu & Toth 2015) in support of ESA-ACES, geodesy and other proposed missions. As indicated in these references, the required level of analysis for accurate determination of time and position for the type of orbit considered here with velocities of ~ 4 km/s, the relativistic effects to fourth order in $[1/c]^4$ for both potential and velocity will be required.

A potential opportunity for this mission is to observe and measure small Post-Newtonian (PN) effects on the satellite orbit (Iorio 2019). The post-Newtonian approximation is used to solve Einstein's field equations for cases in which motions are slow compared to the speed of light and gravitational fields are weak. This approximation is helpful for interpreting experimental tests of general relativity (Will 2014). The high-performance two-way Doppler-cancelling laser links will provide unprecedented stability and accuracy on the satellite velocity-projection (range-rate) and range between the ground stations and the satellite. Combining that with knowledge of the gravity field can give an unprecedented accuracy of the satellite orbit via an approach similar to that used to analyze the orbits of geodetic satellites. Additionally, the velocity and range will have unprecedented precision. An onboard accelerometer can accurately measure non-gravitational disturbances and augment the orbit solution.

The precise orbit determination (POD) for FOCOS will enable this mission to contribute to the difficult task of measuring the Lense-Thirring precession that results from a distortion of the spacetime metric due to the rotation of a massive object. Combined analysis of the orbits of FOCOS and LAGEOS 1 & II and LARES may help to reduce troubling systematic errors and the uncertainties of the frame dragging of orbits (Ciufolini et al. 2017; Everitt et al. 2015; Iorio 2019; Lucchesi et al. 2019).

The Shapiro Time Delay of GR, caused by the spacetime dilation of signals passing near a massive object, will contribute to the timing and ranging measurements. At the cm and mm level of orbit accuracy, there are other known GR effects that have not yet been observed. These include a small geodetic effect that changes the semimajor axis of the orbit by about 1-3 cm/day (Iorio 2019; Nordtvedt 1995). With POD and removal of non-gravitational perturbations, these predicted effects may be observable.

(Wolf & Blanchet 2016) point out that EEP violating terms of order $GM/c^2$ due to the Sun and Moon can be constrained if a satellite clock is compared to two ground clocks separated by intercontinental distances (Altschul et al. 2015; Blanchet et al. 2001; Ciufolini et al. 2017; Iorio 2019; Müller, Soffel & Klioner 2008; Petit & Wolf 2005). This will be part of the baseline mission of FOCOS.

- **Dark Matter searches in Space and on Earth (Standard Model/LPI):**

The coupling of ultra-light scalar bosonic dark matter to the Standard Model can lead to temporal and spatial changes of the values of fundamental constants and thereby the frequencies of atomic transitions. As a result, high performance clocks can search for dark matter signatures in a variety of ways (Arvanitaki, Huang & Van Tilburg 2015; Derevianko & Pospelov 2014; Safronova et al. 2018). In one example, a collaboration created a world-wide network of atomic clocks to look for nearly simultaneous fluctuations in their clock-cavity systems that could indicate passage of a dark matter discontinuity (Wcisło et al. 2018). New limits on such transient effects have been recently reported using a European network of fiber-linked optical atomic clocks (Roberts et al. 2020). Additionally, new approaches for dark matter searches with clocks are being devised (see for example, (Bergé et al. 2019; Hees et al. 2018; Kalaydzhyan & Yu 2017; Leefer et al. 2016; Savalle et al. 2019b; Stadnik 2020)), and some of these can be best implemented with space platforms to minimize the back-action of ordinary matter on the dark matter scalar fields. For example, (Stadnik 2020) pointed out that, for certain interactions and signs of the coupling constants, the interaction of dark matter field with regular matter makes the scalar-field amplitude dependent on the local matter density, leading to the spatial dependence of the fundamental constants in the vicinity Earth. The resulting static signatures (Stadnik 2020) produce stronger bounds (for certain signs of the assumed coupling) of new fields than do transient searches (Roberts et al. 2020; Wcisło et al. 2018), and comparing ground and space clocks can significantly improve the bounds on such ultralight fields. By directly linking higher performance clocks at the sub-$10^{-18}$ level, the FOCOS mission would help establish a high precision clock network of ground clocks and the space clock for dark matter searches with 100-1000x higher sensitivity. Thus, the FOCOS mission is likely to be well placed to respond to new possibilities for the dark matter detection, not only avoiding earth screening, but also taking advantage of its longer baseline to extend dark matter searches to longer-wavelength, lower-mass dark matter objects.

Related, the long baseline of FOCOS offers a clear advantage for searches of bursts of exotic low-mass fields (ELFs) emitted during powerful astrophysical events, such as black hole mergers (Dailey et al. 2021). ELFs can be emitted due to a variety of scenarios, including the stripping of clouds of dark matter fields surrounding merging black holes. ELFs can be caused by mergers of black hole singularities when the effects of quantum gravity are anticipated to be of crucial importance. Arrival of ELFs bursts in the solar system would be delayed with respect to gravitational wave bursts, with GW observatories providing a time trigger and a sky location of the progenitor. Compared to dark matter constituents moving at 300 km/s galactic velocities, ELFs move at nearly the speed of light, requiring a much larger baseline to track the leading edge of the ELF burst.

- **Fundamental physics tests (LPI) by a global clock network:**

  Connecting clocks world-wide would also enable new applications in fundamental and applied science. For instance, an interesting theme in fundamental physics research is that our fundamental constants may not be constant throughout space or in time (Dirac 1938; Martins 2017). Such an inconstancy violates Lorentz Position Invariance and the Universality of Free Fall; indeed, many extension theories predict such a violation. The most sensitive tests to date have compared atomic spectra, either between Earth and quasar-absorption spectra in high-redshift gas clouds, or between atomic clocks. While some astronomical spectral comparisons have hinted at possible variations (Webb et al. 2011), ground based measurements have shown no present-day variations in $\alpha_{FS}$, the fine structure constant, or $\mu$, the proton-electron mass ratios, at the $10^{-18}$/year level (Lange et al. 2021). So far, such ground-based tests have been limited to comparisons between clocks in the same laboratory or those connected by direct fiber links. Connections via FOCOS links would enable fundamental tests between larger numbers of high-performance clocks and provide checks for non-null results, as well as enabling new tests between remotely located systems with larger differential sensitivities to drifts.

- **Worldwide time at the 100-fs level:**

For international time and frequency distribution, the FOCOS clock would provide an international time reference with timing stability at the 100 femtosecond level and with a frequency instability of < $1 \times 10^{-18}$, without interference from the $3 \times 10^{-16}$ tidal fluctuations and $1 \times 10^{-17}$ gravitational noise on the Earth's surface. This mission would lay the groundwork for a future upgrades and augmentation to GNSS systems based on a network of satellite-borne optical clocks with optical links to ground and additional cross-links between satellites (Berceau et al. 2016; Schuldt et al. 2021). Such a system would generate an improved coordinate time reference, analogous to UTC for earth and space (e.g. GNSS) applications but with performance at the 100-fs level, compared to the current nanosecond level .

- **Precision Geodetic Referencing (mm-level):**

    Given that a 1 cm altitude change on earth leads to a $1 \times 10^{-18}$ clock frequency change due to the gravitational redshift, the FOCOS clock could enable mm-level geodesy when a sufficiently accurate clock is used as the earth reference (Mehlstäubler et al. 2018). The FOCOS reference with two-way Doppler-cancelling laser links will provide high accuracy range determination between the satellite and compact, transportable ground terminals. That range data can augment and verify spatial locations for geodetic referencing and Terrestrial Reference Frames (TRF). It will have a higher precision than alternative methods and can complement existing systems based on GNSS, VLBI, DORIS and ILRS and geodetic satellites (Blewitt & Bohm 2019; Ciufolini et al. 2017; Iorio 2019; *LAGEOS-I (Laser Geodynamics Satellite-I)/LAGEOS-II*; Lucchesi et al. 2019; National Academies of Sciences 2020). The two-way laser links should provide range information with a precision better than 1 mm. The system will have the capability to make real time measurements of Earth tides (from range) and perhaps even microseisms (from Doppler). In addition to the general application in geodesy, the high accuracy clock in space can facilitate long distance leveling across mountain ranges and continents by comparing clock frequencies at two locations via the space clock asset. Furthermore, accurate clocks passing close to the Earth can be used to directly measure the gravity potential, providing an alternative method for time-varying gravity measurement of the earth due to slow mass movements (Tapley et al. 2004) .

## 3. Mission Description

### A. Mission Concept

The basic idea for the FOCOS mission is to compare the frequency and time of a stable and accurate clock in space with those of earthbound clocks via an accurate earth-to-space time/frequency laser link. In this mission the onboard clock will experience a continually varying gravitational potential as the spacecraft moves through its eccentric orbit (see Figure 4), continually shifting its frequency relative to that of clocks at earth ground stations. Figure 4(b) shows the frequency variation of a spaceborne clock throughout an eccentric orbit due to the gravitational redshift. For this experiment the measurable will then be the difference between the minimum clock frequency, at periapsis, and the maximum, at apoapsis. Repeated measurements over many orbits give an accurate difference of the space clock's redshifts, after correcting for perturbations such as the Sagnac effect, special relativistic shifts, non-reciprocal corrections related to asynchronous sampling and the point-ahead angle across the two-way link and dispersion effects (Ashby 2003; Bergeron et al. 2019). Comparing the measured and predicted frequency differences will precisely test general relativity. In practice, the measurement would compare the elapsed time (or phases) between the two clocks over successive time intervals, which is the integral of the frequency difference, both to improve sensitivity and to circumvent dropouts of the link due to turbulence or satellite visibility.

When the spacecraft is far from Earth, it will be visible to a larger number of locations on Earth and its clock frequency will vary slowly. Moreover, because the clock spends most of the orbital period far from Earth, there will be adequate averaging times to reach the low $10^{-18}$ range after a small number of orbits. The clock in this way can serve as an accurate space-time reference (Berceau et al. 2016), assuming that its orbit is precisely known and that corrections are applied for both non-gravitational accelerations and higher-order relativistic perturbations.

**B. Spacecraft Orbit Considerations**

The FOCOS mission takes advantage of the large variation in gravitational potential in an eccentric Earth orbit. A modulation of the gravitational potential modulates the clock's frequency, as compared to an earthbound clock, with an accurately known period. This modulation offers two principal metrology benefits: (i) it enables the possibility to take advantage of the stability of the clocks to evaluate gravitational effects at levels beyond the clocks' accuracies, and (ii) it helps to separate gravitational effects from possible drifts in the clocks and link hardware. While this eccentric orbit, and the resulting modulation of the gravitational potential, is required to meet the primary scientific objectives (tests of the gravitational redshift and Lorentz invariance), a circular orbit in MEO is more natural for other scientific objectives, which rely on connecting the FOCOS clock to ground-based clocks around the earth. For example, future state-of-the-art space-time references may benefit from smaller eccentricities, to reduce the variations of relativistic corrections throughout the orbit. Satellite visibility, which sets the duration and frequency of optical links with ground stations that have high performance optical clocks is an important priority for an international space-time reference.

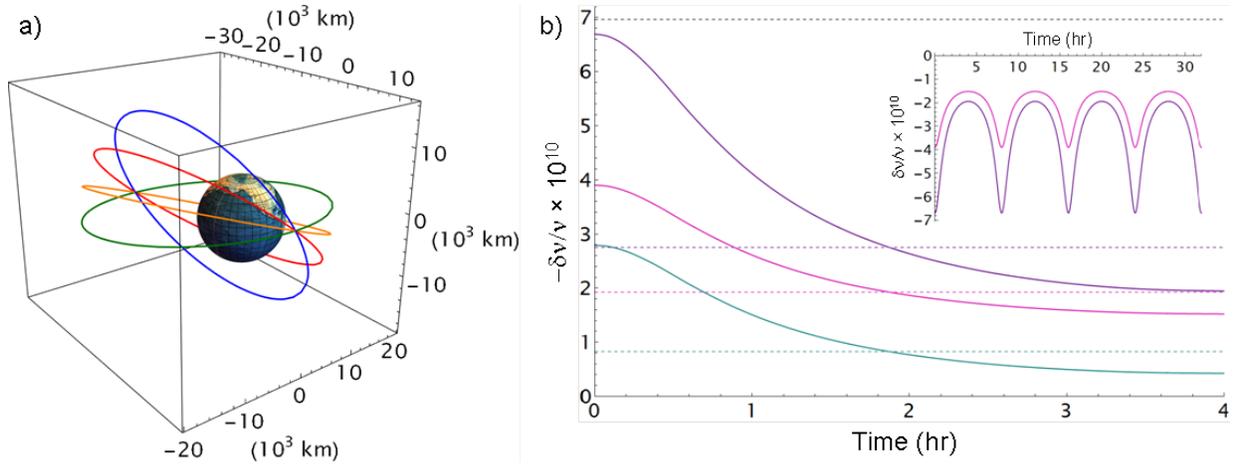

*Figure 4. Mission Schematic: a high-performance optical atomic clock is placed in one of several possible highly eccentric orbits around Earth to modulate the gravitational potential for a sensitive measurement of the gravitational redshift and other relativistic corrections. In addition, a clock in this orbit can serve as an international space-time reference. (a) The orbits (to scale) have an 8-hour period, where the peak elevations of the perigee and apogee are 30 degrees when viewed from 40 N. A tilt of the orbit's minor axis (orange, red, and blue) provides better visibility in both hemispheres. (b) The spaceborne clock's frequency (purple) will vary throughout the orbit because of the summed contributions of the gravitational red shift (magenta) and time dilation (blue-green). This frequency shift, and the corresponding difference in elapsed time, compared to the ground clock, will be measured via low-noise, free space laser links. The black dashed curve is Earth's redshift on the surface, and the colored dashed curves are the corresponding time averages over an orbit. Here we plot the lowest order corrections, and significantly more detailed relativistic calculations are required to realize the accuracy goals of this mission, for example as in (Blanchet et al. 2001).*

Several factors constrain the choice of orbits. To avoid drifts of systematic errors in the clock that would degrade the measurement of the redshift, we favor observing the satellite clock from the same ground location over multiple orbits at both apogee and then at perigee with a minimal gap in time between the two observations, as set by the orbit. A low perigee, which gives a large redshift modulation, limits the visibility at moderate latitudes, such as 40 N, if the apogee is to be visible at the range of a geostationary orbit. Only an 8-hour orbit satisfies these constraints; the satellite clock is observed at perigee, completes 1.5 orbits in 12 hours, and then is observable 12 hours later at apogee after the Earth has rotated 180 degrees. An elliptical 24-hour orbit is also possible, but the apogee range is then large, requiring larger apertures or higher optical powers for the laser link to Earth. A shorter, 18-hour orbit would allow the satellite to be observed at perigee and then 3 hours after apogee, when its range is less than that of a geostationary orbit, but only for a single orbit. An increase in the time gap between observing the perigee and apogee to 36 hours offers a number of solutions, as does 60 hours, but places unfavorably higher demands on the clock's frequency stability. Another possibility is to observe the perigee and the apogee from independent ground stations. This would require chronometric leveling at the level of the accuracy goal for the redshift. We therefore arrive at a baseline orbit with a period of about 8 hours, and a perigee altitude of approximately 5,000 km. For a ground clock located in moderate northern latitudes, tilting the orbital plane so that the perigee is 9 degrees above the equatorial plane allows both the perigee and apogee to have a sufficiently high elevation angle, here taken as at least 20 degrees, for the laser link. As shown in Figure 5, for 40 degrees North, the maximum elevation angle for both perigee and apogee is 30 degrees. The visibility throughout the orbit in both hemispheres further increases if the minor axis of the orbit is significantly tilted with respect to the equatorial plane (e.g. blue and red orbits in Figure 4a and 5). This increased observation time would allow an improved comparison with theory as the frequency shift could be compared through a larger portion of the orbital path, rather than only near periapsis and apoapsis.

The spaceborne clock frequency varies (see Figure 4b) throughout the orbit, and through low-noise, free space laser links, comparisons can be made between the frequencies of the space and ground clocks. The resulting frequency difference tests the redshift of general relativity (red curve), as well as time dilation (blue curve), and higher order relativistic effects. See (Blanchet et al. 2001; Petit & Wolf 1994, 2005) for a more complete discussion of the frequency shifts, including some subtleties related to time and frequency transfer.

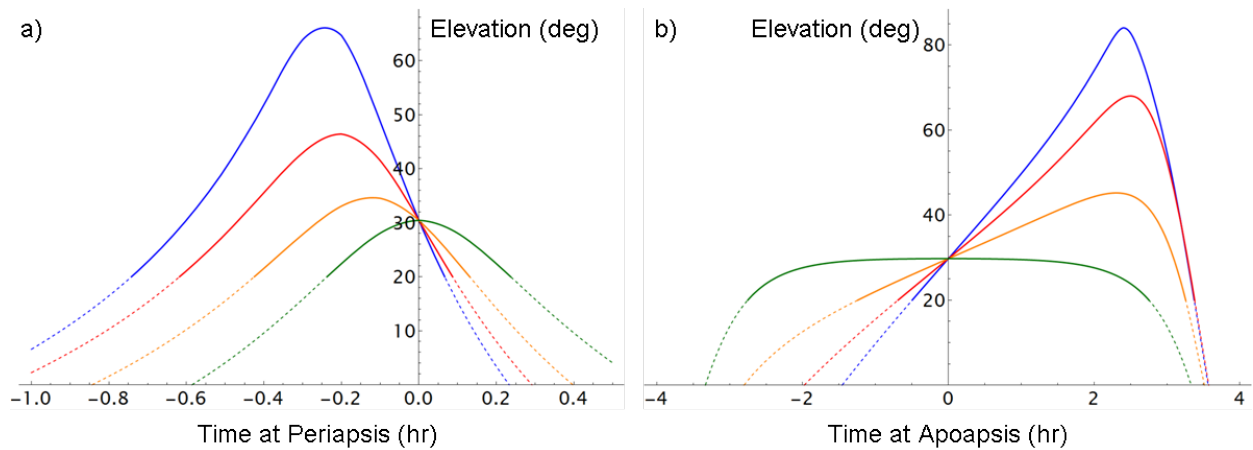

*Fig. 5. Satellite elevation near periapsis and apoapsis for the orbits in Fig. 4a. A tilt of the minor axis of the orbit above the equatorial plane increases the peak elevation and extends the observation time before periapsis. At moderate northern latitudes, a 30 min observation is achievable at periapsis and longer observation times are straightforward at apoapsis. The solid sections of the curves denote elevations above 20 degrees where a laser link from the ground could have a reasonable line-of-sight to the satellite, albeit with ~3 times larger contributions from atmospheric turbulence as compared to a vertical sight path.*

| **Redshift uncertainty budget** | Science Goal 1 ppb |
|---|---|

|  |  | Fractional Frequency δν/ν |
|---|---|---|
| Clock short-term instability |  | $1\times10^{-16}\,\tau^{-\frac{1}{2}}$ |
| Observation time at 5,000 km perigee | 30 min | $2.4 \times 10^{-18}$ |
| Observation time at apogee | 2 hours | $1.2 \times 10^{-18}$ |
| Number of observations | 100 |  |
| Red-shift amplitude |  | $2.4 \times 10^{-10}$ |
| Perigee range uncertainty | 1 mm | $0.7 \times 10^{-19}$ |
| Redshift uncertainty | 1 ppb | $2.4 \times 10^{-19}$ |

*Table 2. Redshift and clock uncertainties for expected clock/time transfer hardware and orbit.*

Given the primary and secondary scientific objectives, we have set the following top-level requirements for the orbit.

- A large modulation of the gravitational redshift, and hence a significant orbit eccentricity, is advantageous to meet the primary scientific objective of a redshift measurement of $10^{-9}$ uncertainty. This requirement is achievable with a clock stability of $1 \times 10^{-16}\,\tau^{-\frac{1}{2}}$, with 100 passes of 30 min of visibility at perigee. Visibility of the satellite at periapsis and at a large distance (potentially apoapsis) is required to measure the difference of the redshift and to validate higher-order relativistic corrections with unprecedented precision.
- To meet the above science objective, we require visibility of the satellite at periapsis and apoapsis for three consecutive orbits at a minimum of one ground station. This will allow more than nine hours of link access. With a projected clock instability of $1 \times 10^{-16}\,\tau^{-1/2}$, nine hours is three times the averaging time required to reach clock accuracy of $1 \times 10^{-18}$. To support a highly synchronized worldwide network of clocks, visibility at least every 24 hours for each ground station is required.
- Central to achieving a sub-ppb test of the redshift is an unprecedented knowledge of the spacecraft's location. A complete analysis of the requirements will follow an analysis done for the ACES mission (Duchayne, Mercier & Wolf 2007), at a higher accuracy. The scale for the orbit determination is challenging. Based on Table 2, the altitude of the orbit must be known to the mm-level around periapsis for the gravitational potential to be known at the 10⁻⁹ level. Additionally, the velocity will need to be characterized at the few micrometer/s level for second-order velocity relativistic corrections. Laser links and clock/cavity stability will need to support range-rate uncertainty below a nm/s to achieve the required frequency uncertainty of 10⁻¹⁸, and range uncertainty of below a mm to achieve a time uncertainty below a picosecond. Repeated observations of the clock through multiple positions in its orbit, including from multiple ground stations, will test LLI and enable precise orbit characterization, along with the verification of higher order relativistic corrections (frame dragging, Shapiro time delay).

## C. Optical Atomic Clock Requirements

The success of this mission depends critically on the technological readiness and performance of the optical atomic clock system. From the scientific objectives, e.g., a redshift uncertainty of $1 \times 10^{-9}$, and an evaluation of technical feasibility, the top-level requirements for the optical atomic clock are:

- Frequency instability noise floor below $2.4 \times 10^{-19}$, consistent with an uncertainty in the redshift modulation of $10^{-19}$, and clock uncertainty of $1 \times 10^{-18}$. Since the redshift will be observed over a 12-hour period, in principle the clock needs to be stable only for this time interval for the redshift measurement. However, consistency over several days, and longer at a lower level, will greatly enhance the averaging over weeks, months, and years. Moreover, this level of uncertainty will be needed to support goals that involve linking international ground-based clocks. We note that constant unknown uncertainties, such as the contribution to the blackbody frequency shift from the uncertainty of atomic transition matrix elements, will not limit the clock comparisons for the redshift measurements or for comparisons of ground clocks.
- Frequency instability of $1 \times 10^{-16} \tau^{-1/2}$. An observation time of 0.5 hrs around periapsis requires an instability of $1 \times 10^{-16} \tau^{-1/2}$ to reach a fractional frequency stability of $2.4 \times 10^{-19}$ after averaging for 100 passes.
- Phase coherence, here defined as an absence of phase slips, for > 12 hours. While some of the science goals will consist of frequency comparisons that in principle are not degraded by occasional phase hops, this is only true if the link is operating continuously. Phase coherence during periods of limited visibility (e.g., near periapsis, cloudy weather, or during high turbulence) will benefit both the measurements and orbit tracking (Meynadier et al. 2018).
- Estimated size, weight, and power (SWaP) of 1.0 m³, 250 kg, and 1 kW, to be compatible with suitable spacecraft host constraints. While the highest performance optical atomic clocks fill multiple optical tables, demonstrated transportable versions suggest that these SWaP levels are feasible.

## D. Optical Time-Frequency Transfer Requirements

The second main mission challenge is the construction of the time/frequency laser link to connect the space-based clock to ground-based clocks. The performance of present-day intercontinental time-frequency transfer based on GPS or two-way microwave links is about 1000x less precise than the requirements for the FOCOS mission. Indeed, current state-of-the-art optical atomic clocks can be precisely compared only if they are geographically close enough to establish a dedicated fiber-optic link, leaving many of the world's highest performing clocks effectively isolated. Figure 6 compares the instability of the proposed optical atomic clock, conventional time-frequency transfer, and the optically based time-frequency transfer.

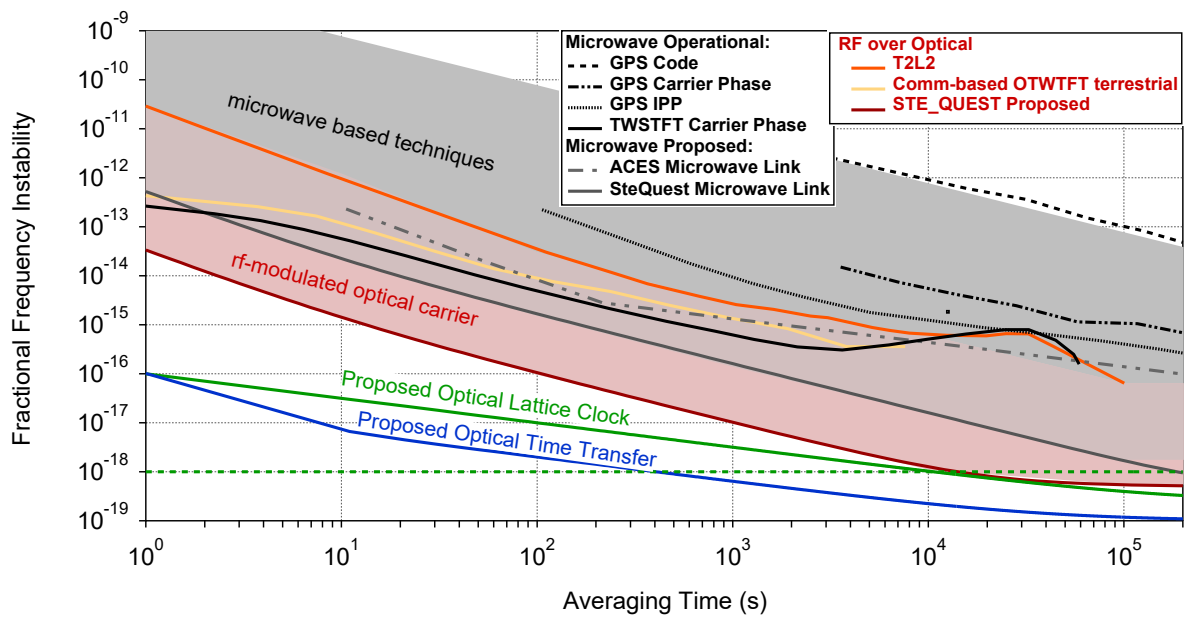

*Figure 6: Fractional frequency uncertainty of the time-frequency transfer versus averaging time, τ, as measured by the modified Allan Deviation.. The grey shaded region encompasses the microwave (or radio frequency) based techniques including GPS (Bauch et al. 2006; Lombardi et al. 2001; Petit et al. 2015), two-way satellite time-frequency transfer (TWSTF) (Fujieda et al. 2014) and the proposed microwave link (MWL) for ACES and STE_QUEST (Delva et al. 2012; Meynadier et al. 2018; STE_Quest Team 2013). The red shaded region encompasses techniques that use an optical carrier but rf-based timing, such as the T2L2 (Samain et al. 2006) and recent MICIUS demonstration (Dai et al. 2020), a communication-based optical two-way time-frequency transfer (O-TWTFT) (Khader et al. 2018), and proposed STE-QUEST link (STE_Quest Team 2013). The solid blue line is the nominal residual instability of the proposed optical time transfer, which follows $10^{-16}/\tau^{3/2}$ reaching a floor of $\sim 1\times 10^{-19}$. The actual performance will be affected by turbulence and orbit visibility. This level of optical time transfer performance has been demonstrated across terrestrial links at terrestrial velocities in a number of demonstrations (Bergeron et al. 2019; Bodine et al. 2020; Deschênes et al. 2016; Dix-Matthews et al. 2021; Shen et al. 2021; Sinclair et al. 2016, 2018), but not from ground to space. The solid green line is the nominal fractional frequency instability, as measured by the total Allan Deviation, for the proposed optical lattice clock, which follows $10^{-16}/\tau^{1/2}$ until it reaches its instability floor of $\sim 2.4\times 10^{-19}$. Its nominal absolute accuracy of $\sim 1\times 10^{-18}$ is shown as the dashed green line.*

The low residual instability of the optical two-way time-frequency transfer must be reached despite frequent and random drop-outs of the link due to turbulence, clouds, weather, airborne debris, as well as the loss of visibility of the satellite during its orbit. This link intermittency leads to requirements for acquisition and tracking of the signal in physical space, in frequency, and in time or phase.

In particular, the strong link intermittency leads to a requirement that the clock comparison approach be based on a phase/time comparison, rather than continuous frequency tracking of the two clocks, as with fiber links. Absolute phase or time transfer between clocks does not require a continuous link; instead, we can simply measure the relative clock phase at two times, $\Delta\theta(t_0)$ and $\Delta\theta(t_1)$. From these two values we can know the clocks' relative time and their relative frequency as $2\pi\Delta f = [\Delta\theta(t_1) - \Delta\theta(t_0)] /(t_1-t_0)$, and do not require any measurements in the intervening times between $t_1$ and $t_0$. The implications of using phase-time comparison as opposed to a frequency comparison are significant, as illustrated by a simple numerical example. Consider two 1-s measurements centered at $t_1$ and $t_0$, separated by 1 hour, where each measurement has a $10^{-16}$ uncertainty. In a pure frequency comparison, the two measurements add in quadrature yielding a residual fractional frequency uncertainty of $0.7 \times 10^{-16}$. Through a phase/time-based comparison (Meynadier et al. 2018), the residual fractional frequency uncertainty is ~ 1 fs/3600s ~ $3 \times 10^{-19}$, over two orders of magnitude better (of course, the measurement may also be limited by the clock's frequency wander at this level). In addition, time transfer opens up a host of applications that depend on the relative phase of the clocks, such as precision navigation, precise orbit determination, ranging, VLBI or future worldwide timescales.

To support the proposed mission, the optical time-frequency transfer should have a residual noise that is below that of the optical lattice clock, leading to the requirements of $10^{-16}$ instability at one second with a floor of ~$2\times 10^{-19}$. (Note that the instability, as measured by the modified Allan deviation, will fall as $\tau^{-3/2}$ if the system is limited by white noise as show in Figure 6.) Recently, optically based two-way time-frequency transfer (O-TWTFT) has been demonstrated that can meet these requirements by exploiting the high reciprocity of single mode links (Bergeron et al. 2019; Bodine et al. 2020; Deschênes et al. 2016; Dix-Matthews et al. 2021; Shen et al. 2021; Sinclair et al. 2016, 2018). However, thus far, these techniques have only been demonstrated over terrestrial distances, of up to 30 km, and at terrestrial velocities, of up to 25 m/s. A key part of the FOCOS technology development roadmap is demonstrating O-TWTFT at mission-relevant (i.e. ground-to-satellite) ranges and velocities. In particular, there are three main challenges to be met. These are to demonstrate optically based time-frequency transfer at (1) satellite-relevant ranges and (2) velocities, and (3) with a lower SWaP, which all seem feasible.

Key top-level requirements for the optical two-way time-frequency transfer include:

- Frequency bias below $2\times 10^{-19}$.
- Residual instability (modified Allan Deviation) below the atomic clock stability for averaging times of 10-s or longer.
- Ability to operate with high intermittent link availability, which includes the ability to track the absolute time/phase offset between the ground clock and satellite clock with no cycle ambiguity.

- Physical acquisition/tracking of the single-mode link between satellite and ground station that also accounts for the point-ahead angle between the launch and receive beams, as in coherent free-space optical communications.
- Ancillary output of the absolute time-of-flight across the link, from well-defined reference planes in the transceivers, to under a picosecond, to support the geodetic scientific objectives.
- Phase-coherent connection between the ground-to-space optical clock time and co-located GNSS receivers, to connect the system to UTC.

## 4. Mission Implementation

Here we present a mission plan including a proposed orbit, the atomic clock system, and the optical timing link, to meet the requirements presented in Section 3, which were based on the diverse science objectives. The level of detail provided here is intended to show the feasibility and to highlight some of the technical challenges. The FOCOS mission will rely on an extensive technology development program, including a system trade study, system engineering, and field testing.

### A. Mission Orbit

To meet the Science goal of a gravitational redshift test of 1 ppb, we have chosen a nominal Earth orbit with the following parameters, which will be more precisely specified by future detailed studies:

**Perigee:** 5,000 km altitude

**Apogee:** 22,800 km altitude

**Period:** 8 hours

**Redshift:** $2.38 \times 10^{-10}$ observed variation and $2.19 \times 10^{-10}$ average shift

A large variation of the redshift increases the signal size and thus improves the constraints we can put on redshift violations and other gravitational couplings. However, increasing the eccentricity of the orbit shortens the observation time at periapsis or requires a larger apoapsis range (e.g., for a 24-hour orbit), which in turn reduces the signal-to-nose of the time transfer link.

### B. An Optical Atomic Clock Based on One or Two Atomic Samples

An optical atomic clock consists of a laser local oscillator whose frequency is locked to a narrow linewidth transition of an atom or ion. The highest performing clocks use trapped atoms or ion(s) to reduce first-order Doppler contributions to the clock uncertainty and interrogate the atoms or ions with a laser pre-stabilized to a narrow resonance of an optical cavity. This pre-stabilization reduces the laser linewidth to much less than 1 Hz on short-time scales (< 10 s), which enables the high-resolution spectroscopy of high performance clocks. The atomic transition provides long-term frequency stability and, with sufficient control of environmental parameters, a reproducible absolute knowledge of the laser frequency. The highest performing clocks trap clouds of neutral atoms with optical standing waves (i.e., an optical lattice) or ions with radiofrequency traps (Figure 7 shows a schematic of an atomic clock that traps clouds in two separate lattices). The long interaction times provided by trapped atoms have led to optical spectroscopic line widths narrower than 1 Hz, yielding line quality factors (Q's) exceeding $10^{15}$. Frequency instabilities and inaccuracies much more precise than the atomic linewidth, fractionally to as high as parts in $10^{-18}$ or $10^{-19}$, are realized by using large numbers of atoms and/or by averaging the clock's frequency. In this way, the atom-stabilized laser is useful as a frequency reference and as a clock, when connected to a frequency comb that can faithfully translate or down-convert the optical signal to a radio frequency signal to be counted.

For this mission we plan to use an optical lattice clock based on a suitable species of neutral atom, nominally ytterbium. Yb and Sr systems have demonstrated the highest performance for the combination of accuracy and stability, both of which are essential to achieve the FOCOS Science Goals. The high frequency stability of optical lattice clocks results from the large number of atoms (~5000) that are simultaneously probed, effectively averaging the quantum noise of many single atom transitions. The required goal of $1 \times 10^{-16} \tau^{-1/2}$ fractional frequency instability is consistent with state-of-the-art lattice clock performance, which is currently limited mainly by thermal mirror coating noise for the pre-stabilization cavities, and not the number of trapped atoms.

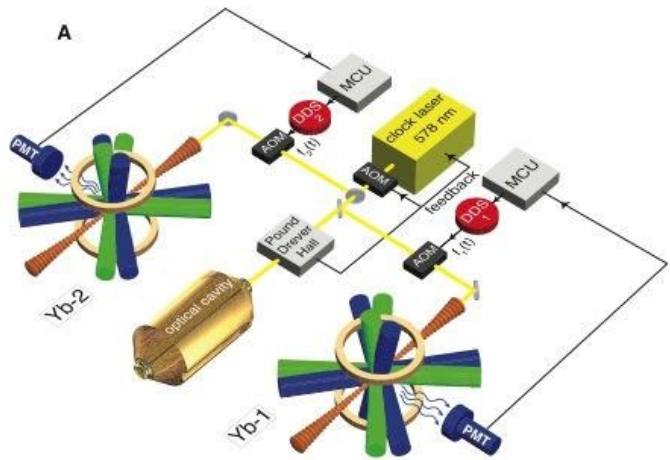

*Figure 7: Schematic of an optical clock based on one laser, one cavity, and two lattice-confined atomic ensembles. Using two ensembles eliminates dead time, greatly reducing the performance requirements of the optical cavity. A more compact design traps two ensembles in two separated lattices in a single vacuum system.*

The highest performing clocks are currently only demonstrated in large laboratory-based research projects. Increasing the technological readiness level of lattice clocks for space deployment (including the lasers and optics, physics package, optical cavities, electronics, and environment control), while meeting the clock performance specifications for this mission, presents significant challenges. The case of the optical reference cavity at first seems particularly daunting. Fluctuations of the length of the cavity, which consists of two mirrors optically contacted to an ultra-low expansion glass spacer, lead directly to frequency fluctuations of the pre-stabilized laser that probes the atoms. The atoms are sensitive to even very fast frequency fluctuations and these significantly degrade the clock stability when the atoms are periodically interrogated. This is the case for most systems, where the atoms are prepared (i.e., trap and cooled) before they are interrogated by the laser (Westergaard, Lodewyck & Lemonde 2010), The most stable clocks therefore isolate their optical cavities so they have extremely low levels of vibrational and thermal noise to reduce this so-called "Dick Effect" noise. While there are numerous designs to mount low-noise cavities for space applications (see for example, (Webster & Gill 2011)), achieving a cavity with a noise floor of $1 \times 10^{-16}$ in space appears challenging.

A different approach alternately probes two (or more) atomic ensembles with the same clock laser (see Figure 7), thereby considerably relaxing the cavity requirements. This eliminates the measurement dead time (Schioppo et al. 2017; Westergaard, Lodewyck & Lemonde 2010) and has allowed clock instabilities to reach the quantum noise limit given by the number of atoms. Here, this technique allows us to relax the cavity requirements so that the cavity-based noise and atom number noise limits are comparable.

| **Instability Budget** | **Single Atomic Ensemble** | **Dual Atomic Ensembles** |
|---|---|---|
| Cavity noise floor | $1 \times 10^{-16}$ | $1 \times 10^{-15}$ |
| $\sigma_{Dick}$ | $7 \times 10^{-17} \tau^{-1/2}$ | $3 \times 10^{-17} \tau^{-1/2}$ |
| $\sigma_{Atom}$ | $2 \times 10^{-17} \tau^{-1/2}$ | $2 \times 10^{-17} \tau^{-1/2}$ |
| $\sigma_{Tech}$ | $3 \times 10^{-17} \tau^{-1/2}$ | $3 \times 10^{-17} \tau^{-1/2}$ |
| $\sigma_{Link}$ | $1 \times 10^{-16} \tau^{-1}$ | $1 \times 10^{-16} \tau^{-1}$ |
| **Total instability** | $8 \times 10^{-17} \tau^{-1/2}$ | $7 \times 10^{-17} \tau^{-1/2}$ |

*Table 3. The sources of frequency instability, σ, for the space-clock and time transfer (as fractional frequency). In each case we consider 3,000 atoms in an atomic sample. The value for the total instability is given for averaging times, τ, longer than 10 s (i.e., after link noise has dropped below $1 \times 10^{-17}$). Note that a comparison with a higher-performance laboratory ground clock would only slightly increase the net measurement instability.*

**Clock Instability:** Table 3 summarizes the instability sources and highlights the difference in cavity requirements for the single- and dual-ensemble approaches. For a single ensemble, a more advanced and challenging cavity supports the desired instability. For a zero-dead time, dual-ensemble system with clouds of 3000 atoms and a 2-Hz spectroscopic linewidth, the Dick effect suppression allows the desired instability with a lower performance cavity. A 2 Hz linewidth requires a cavity with a noise floor of $\sim 1 \times 10^{-15}$, which can be realized in a compact design (< 10 cm) that is compatible with vibration isolation from spacecraft. In either case, well-established techniques can suppress vibration-induced fluctuations and slow drifts of the path length between the laser and the atomic ensembles and laser links ((Ma et al. 1994)). Better cavity performance may also be considered if direct clock-cavity tests of the variation of fundamental constants will improve such tests of fundamental physics.

**Clock Accuracy:** Several advanced experimental techniques are essential to achieve the desired uncertainty of $1 \times 10^{-18}$. For lattice clocks, the systematic frequency shifts of the greatest concern are typically those due to: (i) the lattice light, (ii) stray electric and magnetic fields, and (iii) atom-atom interactions (see Table 4). Of particular concern here is the AC Stark shift due to the blackbody radiation (BBR) field that surrounds the atomic sample. One approach has been to include a cryogenic region in the physics package into which the atoms can be shuttled during the clock spectroscopy period (Ushijima et al. 2015). Room temperature solutions are more attractive for spacecraft and these provide comparable or higher clock accuracy; here the atoms would be surrounded by a metal enclosure whose temperature would be well characterized (Beloy et al. 2014). This approach has reduced fractional blackbody uncertainties to less than $5 \times 10^{-19}$, in addition to simultaneously reducing dc Stark shift uncertainties to less than $1 \times 10^{-19}$. Characterization of the lattice light shifts below the $1 \times 10^{-18}$ level has required detailed evaluations of higher order light shifts (Brown et al. 2017). Protocols now exist to set lattice light intensities and to adequately characterize their light shifts.

**Measurement Uncertainty:** While the full space clock uncertainty budget will set limits for direct clock comparisons with ground clocks, tests of GR using the modulation of the gravitational potential in an eccentric orbit (e.g., the gravitational redshift) will have several common-mode systematics, and therefore these can exceed the limit from the clock's inaccuracy. The desired uncertainty of the measurement of the gravitation redshift of $2.4 \times 10^{-19}$ can be obtained with averaging, where the limit ultimately results from systematic variations associated with the orbit. This high level of frequency precision has previously been demonstrated in lattice clock comparisons (McGrew et al. 2018). To reach this redshift uncertainty, we expect ~ 1 °C temperature control will be required for the vacuum system that surrounds the copper atomic enclosure that controls the BBR shift of the clock and sensors with 10 mK absolute uncertainty for the atomic enclosure (we note that of achieve the BBR uncertainty below would require a modest improvement in the knowledge of the BBR coefficient (McGrew et al. 2018)).

| Accuracy budget | Single Clock |
|---|---|
| Frequency Shift | Fractional Frequency δν/ν |
|  |  |
| Lattice Stark shift | $7 \times 10^{-19}$ |
| Blackbody radiation | $5 \times 10^{-19}$ |
| DC Electric Field | $< 1 \times 10^{-19}$ |
| Zeeman Shifts | $2 \times 10^{-19}$ |
| Background gas collisions | $5 \times 10^{-19}$ |
| Atomic density | $1 \times 10^{-19}$ |
| Total Budget | $1 \times 10^{-18}$ |

*Table 4. Accuracy budget for a spaceborne optical lattice clock. Note the perigee-apogee difference comparison should reach uncertainties below the systematic value (see text) via the clock's high frequency stability and averaging the observed redshift differences over 100 orbits.*

**Clock Physics Package SWaP:** The size, weight and power (SWaP) for the clock physics package is the principal driver for the cost of such a mission, so reducing these parameters are critical for the mission's viability. It has been estimated that for a mission cost at the $300 million threshold, the total payload weight would need to be ~225 kg and power consumption ~ 200 W, well below the values of just transportable optical lattice clocks under development today (450 kg and 750 W) without considering the rest of the payload. Given that the lattice clock is the largest contributor to the mission SWaP, reducing its footprint is an even higher priority than improving clock performance, as today's state-of-the-art systems already approach the required levels.

One strategy would be to find new science approaches that reduce the package while maintaining the performance specifications required by the science goals. In particular optimizing the trade-off between the optical cavity package requirements and the vacuum system containing the lattice-trapped atoms will be critical. For example, while generating (and interrogating) two atomic ensembles increases the size, weight, and power consumption (SWaP) of the atomic physics package, the corresponding relaxed requirements for the cavity isolation might well yield a lower overall SWaP (possibly employing already existing cavity construction/isolation technology). Of course, the final choice between the single- and dual-ensemble approaches will also depend on the relative technical maturity of the cavity and multi-trap hardware packages. We note that advances to oven and pumping technology, as well as miniaturization of the trapping optics, could lead to significant SWaP reductions as well.

Perhaps the more extreme challenge is the reduction in power reduction. A reduced physics package with more efficient oven design or alternative atomic source design would serve as an important first step, but continued improvements in laser technology will be essential. The trends towards chip-scale lasers will need to continue and be extended to cover a wide variety of wavelengths (e.g., for Yb, we need lasers at 399 nm, 556 nm, 578 nm, 759 nm, and 1388 nm) at varying power levels.

There is currently considerable work on all of these technologies (principally in support of transportable lattice clock development) and we anticipate that continued progress will support the following SWaP estimates (overall atomic physics package and supporting electronics): 0.5 m$^3$, 250 kg, and 500 W, with optimism that even lower figures will eventually be attainable.

## C. Optical Time-Frequency Transfer

Turbulence is the most significant complication of optical links between the Earth and space, leading to beam pointing, beam spread, and scintillation. The strength of the turbulence along the path from the ground terminal to the satellite will vary by orders of magnitude depending on the location, time-of-day, weather conditions, elevation angle, and other conditions. Moreover, ground-to-space optical links can operate in the high turbulence regime where effects are difficult to quantify analytically. For free-space optical communications, turbulence-induced scintillation presents significant challenges because of the resulting signal fading – once the received intensity drops below the detection threshold, the communication channel is temporarily lost. Such signal fades occur on any laser link for time transfer (Andrews & Phillips 2005; Caldwell et al. 2020; Fridelance 1997; Robert, Conan & Wolf 2016; Sinclair et al. 2016; Taylor et al. 2020; Taylor, Kahn & Hollberg 2020). In addition to turbulence, cloud cover presents a significant source of signal interruption and must likewise be treated.

To this end, the optical time transfer should: (1) efficiently use photons to maintain a sufficient signal-to-noise ratio, despite turbulence (keeping in mind that the received power threshold for optical time transfer can be considerably lower than that required for an optical gigabit per second link), and (2) use a phase-sensitive time transfer approach that can "ride over" short turbulence- or cloud-induced signal fades. Indeed, because turbulence limits the spatial coherence of the beam and because of the required coupling into single-mode fibers, large aperture telescopes provide little benefit unless adaptive optics are used. In addition, a small aperture telescope reduces the cost of both the space instrument and ground terminals. With these considerations, we select a relatively small space-aperture of 10 cm and a slightly larger ground-aperture of 15-cm.

|  | Perigee (5,000 km @30 deg) | Apogee (22,800 km, 30 deg) |
| --- | --- | --- |
| Redshift (referenced to infinity) | $-3.90 \times 10^{-10}$ | $-1.52 \times 10^{-10}$ |
| Link Loss* | −55 dB | −83 dB |
| Velocity | 4 km/s | 1.8 km/s |
| Doppler shift (1550 nm) | 3 GHz | 1 GHz |
| Doppler slew | 3 MHz/s | 1 MHz/s |

\* Approximate value based on the range to the satellite from a latitude of 40N and corresponding diffractive loss from a 10 cm aperture in space and 15-cm aperture on ground, with an additional 10 dB loss at each terminal (20 dB total) to account for transceiver loss, additional coupling loss into the single mode fiber, and turbulence degradation. The actual link loss will depend strongly on the turbulence conditions and could be higher for some ground terminal locations.

*Table 5: Link parameters.*

The choice of orbit directly sets the requirements on the O-TWTFT operation, as pertains to link loss, aperture, transmit power, frequency (Doppler) shifts, frequency slew rates, and accelerations given in Table 5. The link loss values in Table 5 assume a $1/e^2$ full-width beam at 85% of the aperture, with an additional 10 dB penalty at each terminal to account for transceiver loss, coupling losses in and out of single mode fibers, and beam spread. Detailed modeling of the link loss as outlined in (Andrews & Phillips 2005; Fridelance 1997; Robert, Conan & Wolf 2016) can produce quantitative models for link availability depending on turbulence and weather.

An example O-TWTFT payload is shown in Figure 8. We consider two-way transmission of light from both a modulated CW laser and a frequency comb, where the nominal transmit powers are 0.1 W for the frequency comb and 0.25 W for the CW laser. The comb laser provides sub-$10^{-18}$ frequency transfer and femtosecond-level time transfer, but with 5 ns to 10 ns ambiguity for a 100-MHz to 200-MHz fiber-based frequency comb (Knabe, K. 2020). The CW laser provides a beacon for tracking, a modulated signal for optical communication, a coarse two-way time transfer to remove the 5-nsec ambiguity, and frequency information for signal acquisition. The functions of the CW laser could be supported instead by additional modulation of the frequency comb light, although this would likely require higher power from the frequency comb.

The free-space optical (FSO) terminal is fully gimbaled to provide agile pointing to different ground terminals, depending on cloud cover and user needs, and to decouple the clock orientation from the pointing. It would be possible to have two FSO terminals on board for common view operation, as in the NASA laser communications relay demonstration (Edwards & Israel 2018). However, adding a second FSO terminal also requires another O-TWTFT transceiver. Given the high performance of the on-board clock, one could instead compare to two ground clocks sequentially with a single agile terminal and a high stability on-board clock. As with FSO terminals for coherent optical communication, the FSO terminals require a beacon laser for physical signal acquisition and tip/tilt correction for the atmospheric effect, a tracking gimbal, single-mode fiber input/output, and a compact integration with the rest of the system. Much of the basic design for the FSO terminal could follow that of existing and future satellite free-space optical communication terminals (Cornwell 2015, 2017; Edwards & Israel 2018; Gregory et al. 2012; Hauschildt et al. 2017; Mynaric).

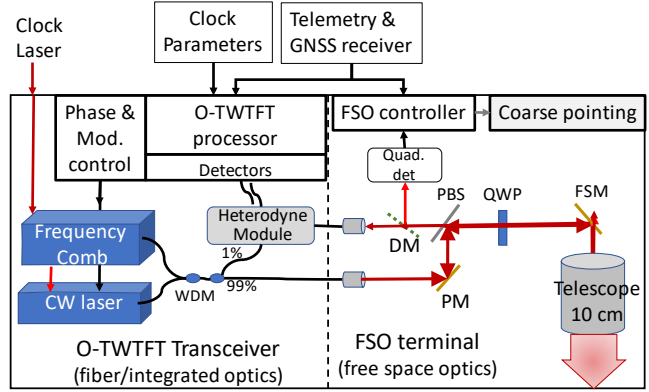

*Figure 8: An optical time-frequency transfer system. O-TWTFT: Optical Two-way Time-Frequency Transfer, FSO: free-space optical, WDM: wavelength division multiplexer, PM: Point-ahead mirror, FSM: fast steering mirror, QWP: quarter-wave plate, PBS: polarizing beam splitter, DM: Dichroic 50% mirror.*

The ground-based FSO terminals share the same requirements with the satellite-based terminals, except they can use a larger aperture. For global time transfer, it is important that the ground terminals do not rely on extremely large ground-based telescopes. Ideally the ground-based FSO terminals would be implemented as relatively compact systems, which can be easily deployed in different locations, such as at national metrology laboratories or locations of interest for geodesy experiments.

We note that there are additional requirements for optical time transfer that do not apply to optical free-space communications. The level of optical phase control and the type of signal processing clearly differ. However, the most critical requirement is that any differential optical path lengths in either the transceiver or the FSO telescope be short and temperature controlled. We define differential optical path lengths as those that are not "common mode" to the output and input light. Fluctuations in these differential optical paths are not removed in the two-way timing comparison. These differential paths include several fiber optic paths within the transceiver as well as the free-space beam paths associated with the point-ahead compensation. For example, a 1 K temperature change of a 1-m optical fiber leads to a 50-fs time shift, the maximum allowed in a system that supports a timescale with 100 fs stability. For reasonable differential path lengths, this corresponds to temperature control requirement at the 0.1 K level.

As noted in the previous section, the size, weight and power (SWaP) is a principal driver for the cost of the mission. There has been recent significant progress towards space-based operation of frequency combs and 10-Watt systems seem viable (Lezius et al. 2016; Manurkar et al. 2018; Sinclair et al. 2014; Timmers et al. 2020). Similarly, there is a significant ongoing effort to develop relatively inexpensive, low-SWAP optical terminals for space-based optical communications in both Europe and the US. As these communication systems move toward higher data rates, they necessarily use coherent processing which has many of the same requirements, e.g. single mode operation, as optical time transfer. As a result, comb-based optical time transfer can leverage much of this technology development in free-space optical terminals for a low SWAP system. Based on current ground-based optical time transfer systems and current LEO or GEO optical communication terminals (Cornwell 2015, 2017; Edwards & Israel 2018; Gregory et al. 2012; Hauschildt et al. 2017; Mynaric), we target a SWaP of 50 liters, 30 kg, and 75 Watts for the O-TWTFT payload.

In addition to the challenges of SWAP reduction, O-TWTFT needs to be extended to satellite-relevant distances and velocities; recent ground-based demonstrations of optical time-frequency transfer with frequency combs have demonstrated the required performance and over turbulent, intermittent links, but only at terrestrial distances of up to 15 km and only at terrestrial velocities of up to 25 m/s. Future satellite-based links will require operation at over 1000x greater distances and 300x greater velocities. Therefore, as with the clock payload, a robust and focused effort will be needed to advance the technology for the proposed mission.

## 5. Final Remarks

Here we have described a space mission that leverages the large recent advances in atomic clocks and time transfer to enable a potential leap forward in tests of General Relativity and the Standard Model, and to provide an international time/frequency reference. While the primary goals described here for this mission focus on tests of fundamental physics (i.e., the gravitational redshift, drifts of fundamental constants, relativistic corrections), a clock orbiting in space would also enable several important secondary applications (i.e., global frequency reference, geodesy reference). Certainly other types of orbits (MEO, GEO, moon, and beyond) and missions can be envisioned that would leverage the clock performance for other types of measurements or applications.

To reach the desired level of readiness on a ten-year time scale, it will be necessary to make significant progress in three principal areas: optical clocks, time transfer, and technological readiness of the various subsystems. Optical clock progress would focus more on reducing the SWaP to reduce mission costs, while increasing the robustness of the physics package and supporting subsystems for long-term remote operation. For time transfer, it will be critical to extend existing time/frequency transfer techniques to satellite-relevant distances and velocities; recent ground-based demonstrations of optical time-frequency transfer with frequency combs have demonstrated the required performance and over turbulent, intermittent links, but only at terrestrial distances and velocities. We emphasize that an aggressive ground development program would more than likely pay for itself in reduced costs for the actual mission.

Over the coming years, we anticipate a series of demonstrations between portable ground-based clocks and airborne clocks (e.g., with airplanes or near earth satellites) over ever-increasing distances. These, in combination with simultaneous reduction in overall SWaP and increases in technological robustness will lead the way to the space instrument. The successful operation of an optical clock in space would not only address the science goals discussed in this paper, but it would also pave the way for a variety of future space missions based on clocks or atom interferometers.

**Acknowledgments**

We thank Alan Kostelecký, Gilad Perez, Yevgeny Stadnik, Peter Graham, and Michael Taylor for helpful comments and discussions. Part of the research was carried out at the Jet Propulsion

Laboratory, California Institute of Technology, under a contract with the National Aeronautics and Space Administration (80NM0018D0004). We also acknowledge support from the NASA BPS Fundamental Physics Program (KG, LH, NN, and CO) and NSF Grant No. OMA-2016244 (MSS).